# Inference of Mixed Graphical Models for Dichotomous Phenotypes using Markov Random Field Model

Jaehyun Park[1], Sungho Won[1,2,3,4]

[1] Interdisciplinary Program in Bioinformatics, College of Natural Sciences, Seoul National University, Seoul, South Korea

[2] Department of Public Health Sciences, Seoul National University, Seoul, South Korea

[3] Institute of Health and Environment, Seoul National University, Seoul, South Korea

[4] RexSoft Corp, Seoul, South Korea

*Corresponding author: Sungho Won

sunghow@gmail.com

+82-880-2714

**Summary:** In this article, we propose a new method named fused mixed graphical model (FMGM), which can infer network structures for dichotomous phenotypes. We assumed that the interplay of different omics markers is associated with disease status and proposed an FMGM-based method to detect the associated omics marker network difference. The statistical models of the networks were based on a pairwise Markov random field model, and penalty functions were added to minimize the effect of sparseness in the networks. The fast proximal gradient method (PGM) was used to optimize the target function. Method validity was measured using synthetic datasets that simulate power-law network structures, and it was found that FMGM showed superior performance, especially in terms of F1 scores, compared with the previous method inferring the networks sequentially (0.392 and 0.546). FMGM performed better not only in identifying the differences (0.217 and 0.410) but also in identifying the networks (0.492 and 0.572). The proposed method was applied to multi-omics profiles of 6-month-old infants with and without atopic dermatitis (AD), and different correlations were found between the abundance of microbial genes related to carotenoid biosynthesis and RNA degradation according to disease status, suggesting the importance of metabolism related to oxidative stress and microbial RNA balance.

# 1 Introduction

Complex diseases can be associated with the underlying network among biological omics data, including the whole genomic DNA profile or whole proteomics (Hawe et al., 2019). Graphical models are widely used to model network structures. The model parameters represent the correlation between each pair of variables conditioned on others and can be expressed as edges in the network. Several methods have been developed to infer the underlying correlations from large datasets in the form of graphs (Hawe et al., 2019). Parametric and non-parametric methods are used and the former include precision matrices of Gaussian models, such as graphical LASSO (Friedman et al., 2008) and joint graphical LASSO (Danaher et al., 2014). Some of these methods, such as GRaFo (Fellinghauer et al., 2013) and GRNBoost (Aibar et al., 2017), utilize non-parametric frameworks, such as random forests (Breiman, 2001).

Previously proposed methods for this purpose have advantages and are widely used. Most studies focused on inferring the network structure without considering disease status. However, biological omics profiles and their networks are expected to remain the same, and certain differences may be responsible for disease status. Gene-gene or protein-protein interactions, such as epistasis (Mackay, 2014), can be an instance of the interplay between biological variables that can be related to phenotypes of interest. For example, a previous study reported that *APOE* and *TOMM40*, whose SNP loci were significantly related to late-onset Alzheimer's disease risk in the Russian population, are correlated at the SNP, gene, and protein levels (Bocharova et al., 2021). In such a scenario, a separate estimation of the networks for patients or controls and their comparison is necessary. Moreover, most interplay is expected to be the same between the cases and controls, and the similarity between them needs to be considered.

The pairwise Markov random field model is a probabilistic model for graphs. Lee and Hastie (2013) suggested a likelihood function for a mixed data pairwise graphical model combining a multivariate Gaussian model and discrete pairwise Markov random field, which can be simplified to either the former or latter if numeric or categorical data are uniquely present, respectively. The conditional dependencies between variables are parameterized into the edge weights in the likelihood function, and the weight nullity corresponds to the conditional independence and lack of edges in the graph. The main bottleneck of this model is that the likelihood function requires multidimensional integration, which is difficult to calculate in practice. Lee and Hastie (2013) overcame this problem by using a product of fully conditioned likelihoods or pseudolikelihood (Besag, 1974; Besag, 1975). The conditional likelihoods for the numerical and categorical variables become linear regression and multinomial distributions, respectively.

Sparsity is generally considered in network inference. The main reasons for this include (1) handling a larger number of features than the sample size and (2) shrinking uninformative edges from the resulting model to zero. In the first proposal, a penalty function composed of matrix norms with a single penalization parameter is used for sparsity. This method does not reflect the possible heterogeneity between types of variables connected to each edge; therefore, Sedgewick et al. (2016) extended the framework and used a penalty function that uses different penalization parameters according to edge type (continuous-continuous, continuous-discrete, discrete-discrete). These methods can successfully model mixed-type data parametrically, but they do not consider the disease status and model only a single network.

Here, the main purpose was to develop a method named fused mixed graphical model (FMGM) with mixed data that can be used for detecting the network associated with disease status and to extend previous methods so that the new method can be used to detect networks associated with disease status. This method extends the framework proposed by Sedgewick et al. (2016), which applies different regularization amounts to each type of edge. It assumes that the difference in networks between cases and controls is sparse and that sparsity is adjusted by incorporating a similar penalty function suggested for the joint graphical LASSO (Danaher et al., 2014).

## 2 Methods

*2.1 Notations and Model Formulation*

We assumed that there were two different groups and our method was designed to infer the network structures of both groups. The two groups included the disease status and exposure status. We assumed $N_m$ independent subjects in group $m$ and each subject was indexed with $n_m$. The total number of participants was $N = N_1 + N_2$. We assumed $p$ numerical variables and $q$ categorical variables, which were used to infer the networks. The numerical and categorical variable vectors from the observation $n_m$ were denoted respectively as $x^{(n_m)}$ and $y^{(n_m)}$, and the $s$-th numerical and $r$-th categorical variables were denoted as $x_s^{(n_m)}$ and $y_r^{(n_m)}$, respectively. The level number for $y_r^{(n_m)}$ was denoted with $L_r$.

Lee and Hastie (2013) proposed a pairwise Markov random field that was used to model the likelihood function as follows:

$$\log p(x^{(n_m)}, y^{(n_m)}; \Theta^{(m)})$$

$$= \sum_{s=1}^{p}\sum_{t=1}^{p} -\frac{1}{2}\beta_{st}^{(\Theta^{(m)})} x_s^{(n_m)} x_t^{(n_m)} + \sum_{s=1}^{p} \alpha_s^{(\Theta^{(m)})} x_s^{(n_m)}$$

$$+ \sum_{s=1}^{p}\sum_{j=1}^{q} \rho_{sj}^{(\Theta^{(m)})} \left(y_j^{(n_m)}\right) x_s^{(n_m)} + \sum_{j=1}^{q}\sum_{r=1}^{q} \phi_{rj}^{(\Theta^{(m)})} \left(y_r^{(n_m)}, y_j^{(n_m)}\right)$$

$$- \log Z(\Theta^{(m)})$$

where $\Theta^{(m)} = \left(\alpha_s^{(\Theta^{(m)})}, \beta_{st}^{(\Theta^{(m)})}, \rho_{sj}^{(\Theta^{(m)})}, \text{vec}\left(\phi_{rj}^{(\Theta^{(m)})}\right)'\right)'$ denotes a vector of parameters that models the network in group $m$, and vec indicates the vec operator. $\beta_{st}^{(\Theta^{(m)})}$ is the edge weight between the $s$-th and the $t$-th numerical variable nodes, $\alpha_s^{(\Theta^{(m)})}$ corresponds to the node potential of the $s$-th numerical variable, $\rho_{sj}^{(\Theta^{(m)})}$ denotes the correlation between the $s$-th numerical and the $j$-th categorical variables, and $\phi_{rj}^{(\Theta^{(m)})}$ denotes the edge weights between the $r$-th and the $j$-th categorical variable nodes. $\rho_{sj}^{(\Theta^{(m)})}$ is an $L_j$ dimensional vector of length with the $l$-th element $\rho_{sj}^{(\Theta^{(m)})}(l)$, and indicates the correlation between variables $s$ and $j$. If all elements are zero, they are conditionally independent. Likewise, $\phi_{rj}^{(\Theta^{(m)})}$ is an $L_r \times L_j$ matrix with the $(l,k)$-th element $\phi_{rj}^{(\Theta^{(m)})}(l,k)$, and variables $r$ and $j$ are conditionally independent if all elements are 0.

Directly maximizing the likelihood function is difficult because the calculation of the partition function $Z(\Theta^{(m)})$ requires multidimensional integration. Alternatively, the likelihood could be replaced with a pseudolikelihood function, the conditional distribution function product for each variable (Besag, 1974; Besag, 1975). Minimizing the pseudolikelihood function is computationally efficient and provides a consistent estimator. The negative log of the pseudolikelihood equation was formulated as follows:

$$f_m(\Theta^{(m)}) = -\frac{1}{N} \sum_{n_m} \left( \sum_{s=1}^{p} \log p\left(x_s^{(n_m)} \big| x_{\backslash s}^{(n_m)}, y^{(n_m)}; \Theta^{(m)}\right) \right.$$

$$\left. + \sum_{r=1}^{q} \log p\left(y_r^{(n_m)} \big| x^{(n_m)}, y_{\backslash r}^{(n_m)}; \Theta^{(m)}\right) \right)$$

$$f(\Theta) = \sum_{m=1}^{2} f_m(\Theta^{(m)})$$

The function was divided by the number of observations to make the pseudolikelihood log scale and the penalty function equivalent. For continuous variables, their conditional distribution was assumed to follow a Gaussian distribution, with the mean in the form of a linear regression model and variance equal to the $\beta_{ss}^{(\Theta^{(m)})}$ reciprocal. For categorical variables, a multinomial distribution with $L_r$ levels was considered for conditional distributions. Lee and Hastie had shown that the negative pseudolikelihood log is jointly convex if $\beta_{ss}^{(\Theta^{(m)})} > 0$ (2013).

Penalty functions based on vector or matrix norms can be used to induce a network sparsity. Lee and Hastie (2013) added the following terms to the penalty.

$$g(\Theta^{(m)}) = \lambda \left( \sum_{t<s} \left|\beta_{st}^{(\Theta^{(m)})}\right| + \sum_{s,j} \left\|\rho_{sj}^{(\Theta^{(m)})}\right\|_2 + \sum_{r<j} \left\|\phi_{rj}^{(\Theta^{(m)})}\right\|_F \right)$$

$\lambda$ is the regularization parameter, $\|\cdot\|_2$ is the $l_2$-norm of a vector, and $\|\cdot\|_F$ is the Frobenius norm of a matrix. $l_2$- and Frobenius norms were used instead of $l_1$-norms because the corresponding variables are mutually independent if and only if all the elements in the vector or matrix are 0.

Sedgewick et al. (2016) modified the penalty by granting different regularization parameters for different edge types.

$$g(\Theta^{(m)}) = \lambda_{cc} \sum_{t<s} \left|\beta_{st}^{(\Theta^{(m)})}\right| + \lambda_{cd} \sum_{s,j} \left\|\rho_{sj}^{(\Theta^{(m)})}\right\|_2 + \lambda_{dd} \sum_{r<j} \left\|\phi_{rj}^{(\Theta^{(m)})}\right\|_F$$

These penalties induce sparsity to separate networks. To induce the sparsity to the difference as well, penalty terms regarding the differences in the network edges were added. Consequently, the subdifferentiable penalty function was set as follows:

$$g(\Theta) = \lambda_{cc} \sum_m \sum_{t<s} \left|\beta_{st}^{(\Theta^{(m)})}\right| + \lambda_{cd} \sum_m \sum_{s,j} \left\|\rho_{sj}^{(\Theta^{(m)})}\right\|_2 + \lambda_{dd} \sum_m \sum_{r<j} \left\|\phi_{rj}^{(\Theta^{(m)})}\right\|_F$$

$$+\lambda'_{cc} \sum_{t<s} \left|\beta_{st}^{(\Theta^{(1)})} - \beta_{st}^{(\Theta^{(2)})}\right| + \lambda'_{cd} \sum_{s,j} \left\|\rho_{sj}^{(\Theta^{(1)})} - \rho_{sj}^{(\Theta^{(2)})}\right\|_2$$

$$+\lambda'_{dd} \sum_{r<j} \left\|\phi_{rj}^{(\Theta^{(1)})} - \phi_{rj}^{(\Theta^{(2)})}\right\|_F$$

The minimization of the target function, a sum of negative pseudo log-likelihood and penalty function, was accomplished with fast proximal gradient method (PGM) (Beck and Teboulle, 2009; Combettes and Pesquet, 2011). For mathematical details of the optimization procedures, please refer to Appendix A.

*2.2 Determination of Penalization Parameters*

For the determination of the penalization parameters, we used stable edge-specific penalty selection (StEPS) (Sedgewick et al., 2016)), a modification of the stability approach to regularization selection (StARS) (Liu et al., 2010). The model was fitted with subsamples drawn without replacement using a single penalization parameter $\lambda$. For each edge type, the edge instability average for each $\lambda$, which corresponds to the empirical probability of the disagreement on having a non-zero edge at each of the values, was calculated. From the largest $\lambda$ value, the value was reduced until the threshold was reached where the threshold was defined a priori. Mathematical definitions and detailed procedures are provided in Appendix B.

*2.3 Simulated Data Analysis*

The simulation data were generated using previously published methods (Danaher et al., 2014; Sedgewick et al., 2016). Two-class data for 100 variables with 250 observations were generated based on scale-free networks (Bollobás et al., 2003). The detailed procedures of data generation and analysis are described in Appendix C.

*2.4 Code Implementation*

The numerical optimization procedures were implemented using the R programming language (https://www.R-project.org/). The number of edges to be optimized was $O(p^2 + q^2)$; therefore, the computational time would be very long if each edge was estimated individually. Therefore, we used 'bigmemory', 'bigalgebra', and 'biganalytics' packages (Emerson and Kane, 2020; Kane et al., 2013) so that the optimization could be parallelized according to edges.

*2.5 Real Data Analysis: Atopic Dermatitis (AD) Data from the Childhood Origin of Asthma*

*and Allergic Diseases (COCOA) Cohort*

For demonstration, FMGM was applied to the COCOA Cohort data (Yang et al., 2014). Ninety-five children with AD were selected from the cohort, and controls were selected using propensity scores calculated based on age, sex, and feeding type. The data used in this study comprised biological multi-omics data and several clinical covariates for the selected samples. Multi-omics data included gene transcriptome profiles and intestinal microbial profiles in 6-month-old children. This analysis was an extension of a previously published study (Park et al., 2021).

2.5.1 Gene Transcriptome Data Using Microarray

Gene transcriptome data were generated from fecal samples of 199 participants using microarray on GeneChip® Human Gene 2.0 ST Arrays (Thermo Fisher Scientific, Inc., Waltham, MA, USA) by Macrogen, Inc. (Geumcheon-gu, Seoul, Republic of Korea). Colonocytes collected from fecal samples were used for total RNA extraction. Signal intensities of 44,625 probes were generated and normalized using the Robust Multi-chip Average (RMA) method. A total of 30,980 probes were annotated with their corresponding genes and used for downstream analysis.

2.5.2 Microbial Compositional Data by 16S rRNA Gene

Microbial compositional profile information was obtained using 16S rRNA amplicons from the stool samples (Lee et al., 2018). Two groups, including 149 and 48 participants, were included in microbial data generation separately, with 454 pyrosequencing and Illumina MiSeq platforms, respectively. Seventy-six genera were commonly observed on both platforms and used in the analyses.

Quality control was performed separately for each sequencing platform. Genera were removed if (1) the genus read count for all subjects was less than 0.05% of the total read counts or (2) the proportion of subjects with at least one read count for the genus was less than 25% of the total subjects. The criteria by Li et al. (2013) and Bokulich et al. (2012) were used, while the actual cutoff values were determined empirically. After quality control, log values of counts per million (log-CPM) transformation were applied to each subject using the R package edgeR (Robinson et al., 2010). Centering and scaling were conducted to adjust batch effects using sequencing.

2.5.3 Microbial Functional Data with Metagenome Shotgun

Microbial functions were profiled via whole-metagenome sequencing using stool samples (Lee et al., 2018). Similar to 16S rRNA amplicon sequencing, metagenome sequencing was performed for two separate datasets with 58 and 40 subjects. The Nesoni high-throughput sequencing data analysis toolset (ver. 0.127) Nesoni clip tool was used to remove Illumina adapter sequences and sequences with lengths < 150 bp in each pair or Q scores < 20. Moreover, human DNA sequences were removed using BBMap with the reference human genome.

Functional profiles were annotated for sequences using Kyoto Encyclopedia of Genes and Genomes Orthology IDs (KO) with HUMAnN2 utility scripts. After annotation, the reads per kilobase (RPK) values for each pathway were retained. A total of 361 pathways were commonly observed from the two datasets, and for each dataset, pathways with subjects having at least one read count of less than 25% of the total subjects were excluded. The remaining data were log-CPM-transformed with edgeR (Robinson et al., 2010). To alleviate the batch effect, centering and scaling were performed separately before combining.

2.5.4 Clinical Covariates

Clinical covariates for statistical analyses included sex, delivery method, feeding method, and AD family history. MissForest was used to fill in the missing information (Stekhoven, 2013). The clinical covariate summary statistics are shown in Table 3 with comparisons using chi-squared tests performed with the R package coin (Hothorn et al., 2006). None of the covariates showed significant differences between patients and controls (chi-squared test, p > 0.05).

2.5.5 Subject Description

84 subjects had full profiles of all omics information and clinical variables, 38 of which were patients with AD. Microbial composition profiles for 37 and 47 participants were generated using 454 pyrosequencing and Illumina MiSeq, respectively, and functional profiles for 56 and 28 participants were obtained separately.

2.5.6 Feature Selection

Sparse DIABLO with a single component was used for feature selection using the R package mixOmics (Rohart et al., 2017; Singh et al., 2016). The design matrix was set according to the original publication, and the number of variables to be selected from each omics dataset was determined using the internal tuning function (range: 5–50).

2.5.7 Network Inference

With the selected features and clinical variables for all subjects, network inference according to disease status was conducted and FMGM was applied. StEPS was run to determine the regularization parameters with ten values ranging from 0.08 to 0.64 evenly spaced on a log2-scale.

# 3 Results

## 3.1 Simulated Data Analysis

An inference result overview is presented in Figure 1. The accuracy (the sum of true positives and true negatives divided by the total number of edges) was fairly high (0.988 overall) for all cases, as expected by the low true edge number and the nature of the network estimation methods (Table 1). The average F1 score and Matthew's correlation coefficient were 0.546 and 0.548, respectively, with relatively high precision (0.627) but relatively low recall (0.495; Figure 2). This result indicated that FMGM was relatively conservative in estimating the networks and differences. The performance was better in estimating the networks than the difference (F1 score = 0.572 vs. 0.410).

Compared to the previous method, the overall performance was improved (Table 1; F1 scores of 0.546 and 0.392). FMGM showed a better performance in the detection of inter-network differences (F1 scores of 0.410 and 0.217), as expected due to the penalty function inducing the sparsity of the network differences. Surprisingly, FMGM showed better performance also in inferring the network edges (F1 scores of 0.572 and 0.492). The previous method had a higher recall compared to the proposed method (0.587 vs. 0.495), but the crucial decrease in precision (0.344 vs. 0.627) was the main reason for the lower F1 score.

*3.2 Real Data Analysis: AD Data from the COCOA Cohort*

Inference results are shown in Figure 3. Of the 946 pairs, 118 and 115 showed non-zero interactions in controls and patients, respectively, but only ten pairs differed between the two networks (Table 2). Among the differential edges, five showed a difference larger than $10^{-3}$. Two interactions, *LINC01036-MIR4788* and *Veillonella*-ko00311 (penicillin and cephalosporin biosynthesis), were negatively correlated in controls, but these correlations disappeared in the AD group. Other three pairs, *Raoultella-Cronobacter*, ko00906 (carotenoid biosynthesis)-ko03018 (RNA degradation), and ko00906-ko04066 (HIF-1 signaling pathway), were negatively correlated in both groups, but the correlation was weaker in patients with AD. Among these, ko04066 is a pathway related to humans, and the related results might be spurious.

*LINC01036* and *MIR4788* expressions were negatively correlated in controls, which disappeared for patients. None of the genes had reported functions or annotated diseases. Since the functions of many lncRNAs and miRNAs are still unknown, downstream analysis of these genes may provide new insights into atopic eczema functions and pathogenesis.

*Veillonella* was found to be dominant in one of the main enterotypes in breast- and mixed-fed infants, although the prevalence of AD did not seem to be significantly high or low compared with other enterotypes in both groups (Lee et al., 2018). The abundance of *Veillonalla* did not differ by disease status but was negatively associated with microbial genes related to the biosynthesis of antibiotics (ko00311: penicillin and cephalosporin). This relationship disappeared in patients with AD, suggesting that AD pathogenesis might be related to the loss of balance between antibiotics and specific intestinal genera.

Similarly, negative interactions between *Raoultella* and *Cronobacter* spp. were weakened in patients with AD. No genus has been associated with atopic eczema. However, some features of microbes may be related to disease status. A species belonging to the *Raoultella* genus, *R. ornithinolytica*, converts histidine to histamine (Kanki et al., 2002), a molecule that contributes to inflammatory responses. *Cronobacter* is related to diverse healthcare-related infections, such as neonatal meningitis (Holý and Forsythe, 2014). Therefore, a weakened balance between these two genera could affect the health status of the host, although the actual consequences need to be studied.

The number of microbial genes related to RNA degradation (ko03018) was negatively correlated with the genes annotated with carotenoid biosynthesis (ko00906). As previously described, some carotenoids act as antioxidants that reduce oxidative damage caused by reactive oxygen species (Chew and Park, 2004). Since microbes can perform quality control by degrading oxidated RNA via several proteins, such as MutT and PNPase (Seixas et al., 2021), low amounts of carotenoid series could be related to high oxidative stress of microbial RNA molecules, requiring more microbial genes related to RNA degradation. Thus, disturbances in RNA quality control might be related to AD occurrence, and future analyses could focus on the validation of the results.

## 4 Discussion

In this study, FMGM, a method to infer networks from mixed data in two classes, was proposed. The method uses likelihood functions of pairwise Markov random fields and penalty functions that make the networks and their differences sparse. The fast proximal gradient method (PGM) was used to solve the optimization problem, and the internal minimization problem was resolved using a fixed-point approach. The method showed superior performance in terms of lower type 1 errors and better F1 scores compared to inferring networks according to classes separately (Sedgewick et al., 2016). Surprisingly, FMGM showed higher performance not only in inferring the differences but also in inferring the network structures, and this may be due to the complementary property of the inference processes. FMGM requires six regularization parameters to be determined, but the StEPS method requires much less effort. To the best of our knowledge, this is the first parametric approach to the simultaneous inference of networks in multiple classes.

Type 1 errors were alleviated much better in the FMGM than in the previous method. We also found that FMGM exhibited much better performance in inferring the networks themselves. This may be because the penalty function in FMGM makes the resulting networks similar, and a network from one class can be considered as 'reference information' in the inference of a network in another class. Thus, the inference precision may benefit from the optimization complementary properties.

However, despite the FMGM advantages, it has some limitations. First, FMGM is computationally intensive; a single run of FMGM for 500 subjects with 100 variables took 3–4 h (CPU Intel Xeon 12Core 24 threads, RAM DDR4 16G). The most computationally intensive part is the calculation of $p_L(\Lambda)$, which involves the sequential optimization of parameters in the two groups. The parameters in the two classes must be optimized individually until they converge, and this requires a considerable amount of time. Using looser cutoffs for convergence or a smaller iteration number may avoid this problem. However, this may affect the overall inference accuracy, and the cutoff should be set to find the appropriate balance between time consumption and inference accuracy. In our implementation, the cutoff was set empirically to a rooted mean squared deviation (RMSD) of $10^{-5}$, since this cutoff did not show a significantly longer running time but had better target function values compared to looser cutoffs. Another solution is to implement the method in faster computer languages such as C++.

Second, StEPS (Sedgewick et al., 2016) was utilized for regularization parameter selection. StEPS assumes independence of the edges by type, which is violated in the FMGM setting. If the weight of a specific edge is fixed at zero in both classes, the difference also decreases to zero. Because the regularization parameter values are determined without considering the dependence, the values corresponding to the difference can become large, and the actual inference can be more conservative than expected. Because of this dependence, we also set the StEPS instability cutoff to 0.1 and found that the performance was not appreciably different from the original settings (results not shown). Further studies that resolve this violation and develop a parameter selection method that considers dependence should be performed.

In conclusion, we proposed fused MGM (FMGM), a method to infer networks from mixed data in two classes. The method uses likelihood functions of pairwise Markov random field and penalty functions that make the networks and their differences sparse. The fast proximal gradient method (PGM) was used to solve the optimization problem, and the internal minimization problem was resolved using the fixed-point approach. The method showed superior performance in terms of higher precisions and F1-scores compared to inferring networks by classes separately using the previous method by Sedgewick et al. (2016). Future studies should be conducted to improve the method and analysis, such as solving the time consumption problem and the possible StEPS strictness in determining penalization parameters.

## Code Availability

R codes are available at https://github.com/JaehyunParkBiostat/fusedMGM_R_package, and

an example dataset and a script for analyses can be viewed at https://figshare.com/articles/dataset/FMGM_synthetic_data_example_zip/20509113.

# Appendix A. Mathematical Details of Fused Mixed Graphical Model

**Target Function Minimization**

To minimize the target function, the fast PGM was used (Beck and Teboulle, 2009; Combettes and Pesquet, 2011). As suggested by Beck and Teboulle (2009), the backtracking step-size rule and monotonicity were implemented. First, we considered the following terms:

$$F(\Theta) := f(\Theta) + g(\Theta)$$

$$Q_L(\Theta, \Lambda) := f(\Lambda) + \langle \Theta - \Lambda, \nabla f(\Lambda) \rangle + \frac{L}{2} \|\Theta - \Lambda\|^2 + g(\Theta)$$

$$p_L(\Lambda) := \arg\min_{\Theta} \{Q_L(\Theta, \Lambda)\}$$

$$= \arg\min_{\Theta} \left\{ g(\Theta) + \frac{L}{2} \left\| \Theta - \left( \Lambda - \frac{1}{L} \nabla f(\Lambda) \right) \right\|^2 \right\}$$

where $\nabla f(\Lambda)$ denotes the first partial derivative of $f$ at point $\Lambda$, which has closed forms. The updated parameter estimates in the $k$-th step were denoted as $\Theta_{(k)}$, and the initial value was defined as $\Theta_{(0)}$. $L$ is the estimate of the Lipschitz constant of $f$ and its update at the $k$-th step was denoted by $L_{(k)}$. The $L_{(k)}$ update was performed by backtracking, where the smallest non-negative integer $i_{(k)}$ that satisfies the following equation was sought:

$$\bar{L} = \eta^{i_{(k)}} L_{(k-1)}$$

$$F\left(p_{\bar{L}}(\Lambda_{(k)})\right) \leq Q_L\left(p_{\bar{L}}(\Lambda_{(k)}), \Lambda_{(k)}\right).$$

Here, $\eta > 1$ is a multiplier, and $L_{(k)}$ is set as $L_{(k)} = \eta^{i_{(k)}} L_{(k-1)}$. Then, the parameters were updated as follows:

$$K_{(k)} = p_{L_{(k)}}(\Lambda_{(k)})$$

$$t_{(k+1)} = \frac{1 + \sqrt{1 + 4t_{(k)}^2}}{2}$$

$$\Theta_{(k)} = \arg\min\{F(\Theta): \Theta = K_{(k)}, \Theta_{(k-1)}\}$$

$$\Lambda_{(k+1)} = \Theta_{(k)} + \left(\frac{t_{(k)}}{t_{(k+1)}}\right)\left(K_{(k)} - \Theta_{(k)}\right) + \left(\frac{t_{(k)} - 1}{t_{(k+1)}}\right)\left(\Theta_{(k)} - \Theta_{(k-1)}\right).$$

where $t_{(k)}$ determines the weights by which previous estimates are reflected.

The iteration stopped when $\Theta_{(k)}$ converged. The convergence could be determined using (1) the RMSD of all parameter values, or (2) the difference in the target function values. Because the algorithm forced $\Theta_{(k)}$ to not change in some steps to guarantee monotonicity, the number of iterations in a row that satisfied the convergence conditions was set to be larger than 1 (default:3).

In the implementation, the initial value of the estimate of the Lipschitz constant $L_{(0)}$ and multiplier $\eta$ was set to one and two, respectively. Furthermore, to accelerate the procedure, a small positive number $\alpha < 1$ (default:0.9) was multiplied by the estimate in each step before the backtracking step.

**Fixed-Point Approach: Non-diagonal Cases**

To calculate $p_L(\Lambda)$, the function to minimize in group $m$ was written as follows:

$$\frac{L}{2}\left(\beta_{st}^{(\Theta^{(m)})} - \beta_{st}^{(\Lambda^{*(m)})}\right)^2 + \lambda_{cc}\left|\beta_{st}^{(\Theta^{(m)})}\right| + \lambda'_{cc}\left|\beta_{st}^{(\Theta^{(1)})} - \beta_{st}^{(\Theta^{(2)})}\right|$$

$$\frac{L}{2}\sum_{l=1}^{L_j}\left(\rho_{sj}^{(\Theta^{(m)})}(l) - \rho_{sj}^{(\Lambda^{*(m)})}(l)\right)^2 + \lambda_{cd}\left\|\rho_{sj}^{(\Theta^{(m)})}\right\|_2 + \lambda'_{cd}\left\|\rho_{sj}^{(\Theta^{(1)})} - \rho_{sj}^{(\Theta^{(2)})}\right\|_2$$

$$\frac{L}{2}\sum_{l,k}\left(\phi_{rs}^{(\Theta^{(m)})}(l,k) - \phi_{rs}^{(\Lambda^{*(m)})}(l,k)\right)^2 + \lambda_{dd}\left\|\phi_{rs}^{(\Theta^{(m)})}\right\|_F + \lambda'_{dd}\left\|\phi_{rs}^{(\Theta^{(1)})} - \phi_{rs}^{(\Theta^{(2)})}\right\|_F$$

where $\Lambda^* = \Lambda - \frac{1}{L}\nabla f(\Lambda)$. Because the functions included the sum of norm functions, the simultaneous minimization of all parameters might be challenging. Instead, the minimization step concerning the parameters in one group was performed first, followed by minimization with the other group. Minimization according to the group was iterated until the parameters converged.

For the edges between numeric variables, it was relatively easy to solve the minimization problem because the function had the form of a quadratic term and absolute value terms. However, the edges that included categorical variables were too complicated to directly determine the minimization point. To solve this problem, another method similar to Weiszfeld's approach for the Fermat-Weber location problem was used (Weiszfeld, 1937; Kuhn, 1973). Because a similar approach could be applied to $\phi_{rs}^{(\Theta^{(m)})}$, the following statements focused on $\rho_{sj}^{(\Theta^{(m)})}$, the weights of the edges between continuous-discrete variables.

First, a set of non-smooth points $B\left(\rho_{sj}^{(\Theta^{(m)})}\right) := \{0\} \cup \left\{\rho_{sj}^{(\Theta^{(m')})}; m' \neq m\right\}$ was defined. The function values at each point in $B\left(\rho_{sj}^{(\Theta^{(m)})}\right)$ were calculated, and the point with the smallest value was denoted as $\rho_{sj}^{(\min)}$. Second, $\rho_{sj}^{(\min)}$ was determined if it satisfied the following inequation:

$$\lambda'_{cd} \geq \left\| L\left(\rho_{sj}^{(\Lambda^{*(m)})} - \rho_{sj}^{(\min)}\right) + \sum_{\rho^*_{sj} \in B\left(\rho_{sj}^{(\Theta^{(m)})}\right) \setminus \rho_{sj}^{(\min)}} \lambda'_{cd} \frac{\rho^*_{sj} - \rho_{sj}^{(\min)}}{\left\| \rho^*_{sj} - \rho_{sj}^{(\min)} \right\|_2} \right\|_2$$

On the left-hand side, $\lambda'_{cd}$ was replaced with $\lambda_{cd}$ if $\rho_{sj}^{(\min)} = 0$. Similarly, on the right-hand side, $\lambda'_{cd}$ was replaced with $\lambda_{cd}$ when $\rho^*_{sj} = 0$. If the inequality held, $\rho_{sj}^{(\min)}$ was considered the minimization point. This was a modification of the criterion proposed by Katz and Vogl (2010). The derivation of this criterion is described in a later section.

If the inequality was not satisfied, the fixed-point approach was used. Because the target function was convex, the minimization point satisfied the condition that the first derivative was equal to zero at the minimization point.

$$L\left(\rho_{sj}^{\left(\Theta^{(m)}\right)} - \rho_{sj}^{\left(\Lambda^{*(m)}\right)}\right) + \lambda_{cd}\frac{\rho_{sj}^{\left(\Theta^{(m)}\right)}}{\left\|\rho_{sj}^{\left(\Theta^{(m)}\right)}\right\|_2} + \lambda'_{cd}\frac{\rho_{sj}^{\left(\Theta^{(m)}\right)} - \rho_{sj}^{\left(\Theta^{(m')}\right)}}{\left\|\rho_{sj}^{\left(\Theta^{(1)}\right)} - \rho_{sj}^{\left(\Theta^{(2)}\right)}\right\|_2} = 0$$

where $m' = 2$ if $m = 1$ and vice versa. This condition was equivalent to the following statement:

$$\rho_{sj}^{\left(\Theta^{(m)}\right)} = \frac{L\rho_{sj}^{\left(\Lambda^{*(m)}\right)} + \dfrac{\lambda'_{cd}\rho_{sj}^{\left(\Theta^{(m')}\right)}}{\left\|\rho_{sj}^{\left(\Theta^{(1)}\right)} - \rho_{sj}^{\left(\Theta^{(2)}\right)}\right\|_2}}{L + \dfrac{\lambda_{cd}}{\left\|\rho_{sj}^{\left(\Theta^{(m)}\right)}\right\|_2} + \dfrac{\lambda'_{cd}}{\left\|\rho_{sj}^{\left(\Theta^1\right)} - \rho_{sj}^{\left(\Theta^{(2)}\right)}\right\|_2}}$$

If the values of $\rho_{sj}^{\left(\Theta^{(m)}\right)}$ at steps $p = 1,2,...$ were set as $\rho_{sj}^{[p]}$, the $p$-th step of the fixed-point approach was as follows:

$$\rho_{sj}^{[p+1]} = \frac{L\rho_{sj}^{\left(\Lambda^{*(m)}\right)} + \dfrac{\lambda'_{cd}\rho_{sj}^{\left(\Theta^{(m')}\right)}}{\left\|\rho_{sj}^{[p]} - \rho_{sj}^{\left(\Theta^{(m')}\right)}\right\|_2}}{L + \dfrac{\lambda_{cd}}{\left\|\rho_{sj}^{[p]}\right\|_2} + \dfrac{\lambda'_{cd}}{\left\|\rho_{sj}^{[p]} - \rho_{sj}^{\left(\Theta^{(m')}\right)}\right\|_2}}$$

The iteration continued until $\rho_{sj}^{[p]}$ converged. In the implementation, the initial value was set as the weighted sum of the points in $B\left(\rho_{sj}^{\left(\Theta^{(m)}\right)}\right)$ as follows:

$$\rho_{sj}^{[1]} = \frac{\sum_{m' \neq m} \lambda'_{cd}\rho_{sj}^{\left(\Theta^{(m')}\right)}}{\lambda_{cd} + \sum_{m' \neq m} \lambda'_{cd}}$$

**Fixed-Point Approach: Diagonal Cases**

The main drawback of the approach that minimized the function according to the parameters in each group was that the iteration could stop on the non-zero diagonal point, or $\beta_{st}^{\left(\Theta^{(1)}\right)} = \beta_{st}^{\left(\Theta^{(2)}\right)} \neq 0$, even if the point was not a true minimization point. For example, if the values in the function were set as $\beta_{st}^{(\Lambda^{*(1)})} = \beta_{st}^{(\Lambda^{*(2)})} = 0.1$ in $L = 5$, and $\lambda_{cc} = \lambda'_{cc} = 0.8$, the iteration stopped at $\beta_{st}^{(\Theta^{(1)})} = \beta_{st}^{(\Theta^{(2)})} = 0.1$, although the true minimization point was $\beta_{st}^{(\Theta^{(1)})} = \beta_{st}^{(\Theta^{(2)})} = 0$. Therefore, an additional minimization with $\beta_{st}^{(\Theta^{(1)})} = \beta_{st}^{(\Theta^{(2)})}$ fixation was performed if the iteration was halted at the diagonal point. The statements below focus on $\rho_{st}^{(\Theta^{(m)})}$, the edges between continuous variables. However, similar approaches could be applied to $\rho_{sj}^{(\Theta^{(m)})}$ and $\phi_{rs}^{(\Theta^{(m)})}$.

If $\beta_{st}^{(\Theta^{(1)})} = \beta_{st}^{(\Theta^{(2)})}$, the target function was reduced to the following:

$$\frac{L}{2}\left( \left(\beta_{st}^{(\Theta^{(1)})} - \beta_{st}^{(\Lambda^{*(1)})}\right)^2 + \left(\beta_{st}^{(\Theta^{(1)})} - \beta_{st}^{(\Lambda^{*(2)})}\right)^2 \right) + 2\lambda_{cc}\left|\beta_{st}^{(\Theta^{(1)})}\right|$$

This function had a minimization point of zero if the following inequality was satisfied (2010):

$$2\lambda_{cc} \geq L\left|\beta_{st}^{(\Lambda^{*(1)})} + \beta_{st}^{(\Lambda^{*(2)})}\right|$$

The proof for this criterion is provided later in the document. Otherwise, a fixed-point approach was used. The first derivative of the reduced function was as follows:

$$L\left( \left(\beta_{st}^{(\Theta^{(1)})} - \beta_{st}^{(\Lambda^{*(1)})}\right) + \left(\beta_{st}^{(\Theta^{(1)})} - \beta_{st}^{(\Lambda^{*(2)})}\right) \right) + 2\lambda_{cc}\frac{\beta_{st}^{(\Theta^{(1)})}}{\left|\beta_{st}^{(\Theta^{(1)})}\right|} = 0$$

The above equation was equivalent to

$$\beta_{st}^{(\Theta^{(1)})} = \frac{L\dfrac{\beta_{st}^{(\Lambda^{*(1)})} + \beta_{st}^{(\Lambda^{*(2)})}}{2}}{L + \dfrac{\lambda_{cc}}{\left|\beta_{st}^{(\Theta^{(1)})}\right|}}$$

Thus, the $p$-th step was as follows:

$$\beta_{st}^{[p+1]} = \frac{L\dfrac{\beta_{st}^{(\Lambda^{*(1)})} + \beta_{st}^{(\Lambda^{*(2)})}}{2}}{L + \dfrac{\lambda_{cc}}{\left|\beta_{st}^{[p]}\right|}}$$

**Proof of Criterion Judging a Minimization Point**

The proof focused on non-diagonal cases for continuous-discrete edges, and similar approaches could be applied to other cases.

The function to minimize was expressed as follows:

$$f\left(\rho_{sj}^{(\Theta^{(1)})}\right) = \frac{L}{2}\sum_{l=1}^{L_1}\left(\rho_{sj}^{(\Theta^{(1)})}(l) - \rho_{sj}^{(\Lambda^{*(1)})}(l)\right)^2 + \lambda_{cd}\left\|\rho_{sj}^{(\Theta^{(1)})}\right\|_2 + \lambda_{cd}'\left\|\rho_{sj}^{(\Theta^{(1)})} - \rho_{sj}^{(\Theta^{(2)})}\right\|_2$$

$z = \{z(l)\}_{l=1,\ldots,L_1}$ was taken as an arbitrary unit vector, that is, $\|z\|_2 = 1$. If $\rho_{sj}^{(\min)} = 0$, the rate of change at $\rho_{sj}^{(\min)}$ along the direction of $z$ could be calculated as follows:

$$f\left(\rho_{sj}^{(\min)} + tz\right) = f(tz) = \frac{L}{2}\sum_{l=1}^{L_1}\left(tz(l) - \rho_{sj}^{(\Lambda^{*(1)})}(l)\right)^2 + \lambda_{cd}\|tz\|_2 + \lambda_{cd}'\left\|tz - \rho_{sj}^{(\Theta^{(2)})}\right\|_2$$

$$\frac{\partial}{\partial t}f\left(\rho_{sj}^{(\min)} + tz\right)$$

$$= L\sum_{l=1}^{L_1}z(l)\left(tz(l) - \rho_{sj}^{(\Lambda^{*(1)})}(l)\right) + \lambda_{cd} + \lambda_{cd}'\frac{\sum_{l=1}^{L_1}z(l)\left(tz(l) - \rho_{sj}^{(\Theta^{(2)})}(l)\right)}{\left\|tz - \rho_{sj}^{(\Theta^{(2)})}\right\|_2}$$

$$\left.\frac{\partial}{\partial t}f\left(\rho_{sj}^{(min)}+tz\right)\right|_{t\downarrow 0}=L\sum_{l=1}^{L_1}z(l)\left(-\rho_{sj}^{(\Lambda^{*(1)})}(l)\right)+\lambda_{cd}+\lambda_{cd}'\frac{\sum_{l=1}^{L_1}z(l)\left(-\rho_{sj}^{(\Theta^{(2)})}(l)\right)}{\left\|\rho_{sj}^{(\Theta^{(2)})}\right\|_2}$$

$$=\lambda_{cd}-L\sum_{l=1}^{L_1}z(l)\rho_{sj}^{(\Lambda^{*(1)})}(l)-\frac{\sum_{l=1}^{L_1}\lambda_{cd}'z(l)\rho_{sj}^{(\Theta^{(2)})}(l)}{\left\|\rho_{sj}^{(\Theta^{(2)})}\right\|_2}$$

$$=\lambda_{cd}-z\cdot\left(L\rho_{sj}^{(\Lambda^{*(1)})}+\frac{\lambda_{cd}'\rho_{sj}^{(\Theta^{(2)})}(l)}{\left\|\rho_{sj}^{(\Theta^{(2)})}\right\|_2}\right)$$

where $\cdot$ is an inner product operator. Thus, the descent was the greatest in the following direction:

$$R=\frac{L\rho_{sj}^{(\Lambda^{*(1)})}+\frac{\lambda_{cd}'\rho_{sj}^{(\Theta^{(2)})}(l)}{\left\|\rho_{sj}^{(\Theta^{(2)})}\right\|_2}}{\left\|L\rho_{sj}^{(\Lambda^{*(1)})}+\frac{\lambda_{cd}'\rho_{sj}^{(\Theta^{(2)})}(l)}{\left\|\rho_{sj}^{(\Theta^{(2)})}\right\|_2}\right\|_2}$$

If $z$ was replaced with $R$, the change rate was equal to the following:

$$\lambda_{cd}-\left\|L\rho_{sj}^{(\Lambda^{*(1)})}+\frac{\lambda_{cd}'\rho_{sj}^{(\Theta^{(2)})}(l)}{\left\|\rho_{sj}^{(\Theta^{(2)})}\right\|_2}\right\|$$

$\rho_{sj}^{(min)}=0$ was the minimizing point if the descent was non-negative, and this was equivalent to the criterion in Section 2.3, where $\rho_{sj}^{(min)}$ was replaced with a zero vector. Therefore, the criterion was proven in the case where $\rho_{sj}^{(min)}=0$.

Similarly, if $\rho_{sj}^{(min)}=\rho_{sj}^{(\Theta^{(2)})}$, the descent along a unit vector $z$ was derived as follows:

$$f\left(\rho_{sj}^{(\min)} + tz\right) = f\left(\rho_{sj}^{(\Theta^{(2)})} + tz\right)$$

$$= \frac{L}{2}\sum_{l=1}^{L_1}\left(\rho_{sj}^{(\Theta^{(2)})}(l) + tz(l) - \rho_{sj}^{(\Lambda^{*(1)})}(l)\right)^2 + \lambda_{cd}\left\|\rho_{sj}^{(\Theta^{(2)})} + tz\right\|_2 + \lambda'_{cd}\|tz\|_2$$

$$\frac{\partial}{\partial t}f\left(\rho_{sj}^{(\min)} + tz\right)$$

$$= L\sum_{l=1}^{L_1}z(l)\left(\rho_{sj}^{(\Theta^{(2)})}(l) + tz(l) - \rho_{sj}^{(\Lambda^{*(1)})}(l)\right)$$

$$+ \lambda_{cd}\frac{\sum_{l=1}^{L_1}z(l)\left(\rho_{sj}^{(\Theta^{(2)})}(l) + tz(l)\right)}{\left\|\rho_{sj}^{(\Theta^{(2)})} + tz\right\|_2} + \lambda'_{cd}$$

$$\left.\frac{\partial}{\partial t}f\left(\rho_{sj}^{(\min)} + tz\right)\right|_{t\downarrow 0}$$

$$= L\sum_{l=1}^{L_1}z(l)\left(\rho_{sj}^{(\Theta^{(2)})}(l) - \rho_{sj}^{(\Lambda^{*(1)})}(l)\right) + \lambda_{cd}\frac{\sum_{l=1}^{L_1}z(l)\left(\rho_{sj}^{(\Theta^{(2)})}(l)\right)}{\left\|\rho_{sj}^{(\Theta^{(2)})}\right\|_2} + \lambda'_{cd}$$

$$= \lambda_{cd} - z\cdot\left(L\left(\rho_{sj}^{(\Lambda^{*(1)})}(l) - \rho_{sj}^{(\Theta^{(2)})}(l)\right) - \frac{\lambda_{cd}\rho_{sj}^{(\Theta^{(2)})}(l)}{\left\|\rho_{sj}^{(\Theta^{(2)})}\right\|_2}\right)$$

The descent would be the steepest in the following direction:

$$R = \frac{L\left(\rho_{sj}^{(\Lambda^{*(1)})}(l) - \rho_{sj}^{(\Theta^{(2)})}(l)\right) - \frac{\lambda_{cd}\rho_{sj}^{(\Theta^{(2)})}(l)}{\left\|\rho_{sj}^{(\Theta^{(2)})}\right\|_2}}{\left\|L\left(\rho_{sj}^{(\Lambda^{*(1)})}(l) - \rho_{sj}^{(\Theta^{(2)})}(l)\right) - \frac{\lambda_{cd}\rho_{sj}^{(\Theta^{(2)})}(l)}{\left\|\rho_{sj}^{(\Theta^{(2)})}\right\|_2}\right\|_2}$$

If $z$ was replaced with $R$, the descent was reduced to the following:

$$\lambda_{cd} - \left\|L\left(\rho_{sj}^{(\Lambda^{*(1)})}(l) - \rho_{sj}^{(\Theta^{(2)})}(l)\right) - \frac{\lambda_{cd}\rho_{sj}^{(\Theta^{(2)})}(l)}{\left\|\rho_{sj}^{(\Theta^{(2)})}\right\|_2}\right\|_2$$

If the value was non-negative, $\rho_{sj}^{(min)} = \rho_{sj}^{(\Theta^{(2)})}$ was the minimization point. Thus, this criterion also held true in the case of $\rho_{sj}^{(min)} = \rho_{sj}^{(\Theta^{(2)})}$.

**Appendix B. Regularization Parameter Determination with StEPS**

Stable edge-specific penalty selection (StEPS) (Sedgewick et al., 2016), a modification of stability approach to regularization selection (StARS) (Liu et al., 2010), was used for penalization parameter determination. From a dataset of $n$ samples, $N$ subsamples of size $b$ were drawn without replacement. The model was fitted for each subsample using a single penalization parameter $\lambda$. For the edge between variables $s$ and $t$, the fraction of subsamples that contained non-zero edges, $\hat{\theta}_{st}(\lambda)$, was calculated. Edge instability, the empirical probability of disagreement of having a non-zero edge at each $\lambda$ value, was defined as follows:

$$\hat{\xi}_{st}(\lambda) = 2\hat{\theta}_{st}(\lambda)\left(1 - \hat{\theta}_{st}(\lambda)\right)$$

The total instability for each edge type was calculated as follows:

$$\hat{D}_{cc}(\lambda) = \frac{\sum_{cc} \hat{\xi}_{st}(\lambda)}{\binom{p}{2}}$$

$$\hat{D}_{cd}(\lambda) = \frac{\sum_{cd} \hat{\xi}_{st}(\lambda)}{pq}$$

$$\hat{D}_{dd}(\lambda) = \frac{\sum_{dd} \hat{\xi}_{st}(\lambda)}{\binom{q}{2}}$$

The calculated instabilities were monotonized to prevent selecting dense graphs with low instability as follows:

$$\bar{D}_{cc}(\lambda) = \sup_{\lambda \leq t} \hat{D}_{cc}(t)$$

$$\bar{D}_{cd}(\lambda) = \sup_{\lambda \leq t} \hat{D}_{cd}(t)$$

$$\bar{D}_{dd}(\lambda) = \sup_{\lambda \leq t} \hat{D}_{dd}(t)$$

From the largest $\lambda$ value, the value was reduced until the threshold was reached where the threshold was defined a priori.

In the implementation, the number and size of the subsamples in StEPS were determined beforehand. The default values in the implementation were N=20 and b=10√n, as suggested by Liu et al. (2010).

## Appendix C. Data Simulation Details

One hundred variables, 50 of which were normally distributed numeric variables and the others were categorical variables with four levels each, were randomly divided into five equal-size variable sets. Scale-free networks were generated for each of the variable sets based on the method used by Bollobás *et al.* (2003). Briefly, the generation started with only two variables connected, and the edges were iteratively added until all 20 variables in the set were connected. In each step, two non-zero-degree nodes were connected by a probability of 0.3, or a node with 0 degrees was connected to a non-zero-degree node with a probability of 0.7. Non-zero-degree nodes were selected based on their probabilities proportional to their degrees. The directionality and duplicated edges were ignored in the resulting networks. An example of a generated network is shown in Supplementary Figure 1.

The networks from the two classes were simulated. One of the five networks was randomly selected and removed from the first class and the other was selected again and removed from the second class. Each class had four distinct networks, three of them overlapping. For each class, 250 observations were generated using the Markov chain Monte Carlo method. The weight of each edge, $w_{st}$, was randomly sampled from a uniform distribution ranging from 0.5 to 0.8. For each $\beta_{st}$, the value was set to $w_{st}$ and the sign was randomly sampled with even probability. To ensure that the matrix was positive definite, the diagonal elements were set to be the largest among the sums of the absolute values of the edge weights connected to each of the nodes. For $\rho_{st}$, the values were permuted from $[-w_{st}, -.5w_{st}, .5w_{st}, w_{st}]$. For $\phi_{st}$, one parameter in each column and each row was set to be $w_{st}$ and the rest was set to be $-w_{st}$.

The FMGM was run for 50 datasets that were repeatedly simulated using the procedures described above. For regularization parameters, seven values ranging from 0.08 to 0.32, which were equally spaced on a log2 scale, were tested using StEPS with the default parameters. The smallest values with average estimation instabilities below the threshold for each edge type were used.

Additionally, we used a previously published method by Sedgewick et al. (2016) for each class. The causalMGM package is currently unavailable in R, but because FMGM is reduced to causalMGM in one-class cases, the same code implemented in R was used. The same values as the suggested method were used in StEPS, with the same criteria used to determine the penalization parameters.

**Table 1.** Performance measures obtained using the stated network inference methods with the simulated data. Each simulated data contains 100 variables (50 Gaussian and 50 categorical), and 4,950 pairs of variables were subjected to the inference. The average values over 50 repetitions are shown, and the standard deviations are shown in the parenthesis.

|  | Accuracy | Precision | Recall | F1-score | Matthews CC |
|---|---|---|---|---|---|
| CausalMGM for each group |  |  |  |  |  |
|   Overall | 0.974 (0.00908) | 0.344 (0.0987) | 0.587 (0.2233) | 0.392 (0.1381) | 0.413 (0.1289) |
|   Intra-network | 0.980 (0.00666) | 0.508 (0.1536) | 0.591 (0.2265) | 0.492 (0.1726) | 0.504 (0.1509) |
|     Cont-Cont | 0.982 (0.00920) | 0.580 (0.2081) | 0.633 (0.2973) | 0.543 (0.1954) | 0.564 (0.1748) |
|     Cont-Disc | 0.984 (0.00707) | 0.606 (0.1655) | 0.753 (0.2556) | 0.607 (0.1987) | 0.626 (0.1670) |
|     Disc-Disc | 0.969 (0.01792) | 0.207 (0.3244) | 0.211 (0.3516) | 0.201 (0.3378) | 0.190 (0.3431) |
|   Inter-network | 0.963 (0.01529) | 0.148 (0.0687) | 0.573 (0.2242) | 0.217 (0.0846) | 0.267 (0.0982) |
|     Cont-Cont | 0.965 (0.01959) | 0.173 (0.1299) | 0.629 (0.3225) | 0.232 (0.1348) | 0.291 (0.1417) |
|     Cont-Disc | 0.964 (0.01618) | 0.187 (0.1093) | 0.747 (0.2531) | 0.269 (0.0925) | 0.343 (0.1069) |
|     Disc-Disc | 0.959 (0.02423) | 0.060 (0.1222) | 0.194 (0.3598) | 0.090 (0.1788) | 0.092 (0.2096) |
| Fused MGM (FMGM) |  |  |  |  |  |
|   Overall | 0.988 (0.00250) | 0.627 (0.1013) | 0.495 (0.1462) | 0.546 (0.1208) | 0.548 (0.1180) |
|   Intra-network | 0.986 (0.00329) | 0.626 (0.1072) | 0.538 (0.1499) | 0.572 (0.1218) | 0.570 (0.1201) |
|     Cont-Cont | 0.990 (0.00420) | 0.698 (0.1175) | 0.778 (0.1908) | 0.722 (0.1256) | 0.725 (0.1252) |
|     Cont-Disc | 0.991 (0.00259) | 0.947 (0.0325) | 0.570 (0.1702) | 0.696 (0.1326) | 0.722 (0.1089) |
|     Disc-Disc | 0.970 (0.01229) | 0.191 (0.2939) | 0.219 (0.3431) | 0.202 (0.3163) | 0.188 (0.3230) |
|   Inter-network | 0.992 (0.00133) | 0.643 (0.1129) | 0.322 (0.1630) | 0.410 (0.1556) | 0.440 (0.1379) |
|     Cont-Cont | 0.994 (0.00336) | 0.761 (0.2746) | 0.411 (0.3093) | 0.504 (0.2687) | 0.540 (0.2565) |
|     Cont-Disc | 0.993 (0.00176) | 0.784 (0.1470) | 0.374 (0.1703) | 0.489 (0.1767) | 0.527 (0.1567) |
|     Disc-Disc | 0.988 (0.00478) | 0.175 (0.3226) | 0.135 (0.2785) | 0.151 (0.2906) | 0.150 (0.2959) |

**Table 2.** Conditional correlations between variables that indicate differences between 6-month-old patients with atopic dermatitis (AD) and controls.

| Variable 1 | Variable 2 | Interaction in controls | Interaction in patients | Difference |
|---|---|---|---|---|
| *LINC01036* (NR_126347) | *MIR4788* (NR_039951) | -0.117 | 0 | 0.117 |
| *Veillonella* | ko00311: Penicillin and cephalosporin biosynthesis | $-3.562 \times 10^{-3}$ | 0 | $3.562 \times 10^{-3}$ |
| *Raoultella* | *Cronobacter* | -0.676 | -0.507 | 0.168 |
| ko00906: Carotenoid biosynthesis | ko03018: RNA degradation | -0.154 | -0.064 | 0.091 |
| ko00906: Carotenoid biosynthesis | ko04066: HIF-1 signaling pathway | -0.118 | -0.094 | 0.023 |
| Gender | *C16orf72* (NM_014117) | $1.725 \times 10^{-12}$ | 0 | $1.725 \times 10^{-12}$ |
| Gender | *C16orf72* (NM_014117) | $2.177 \times 10^{-12}$ | 0 | $2.177 \times 10^{-12}$ |
| Delivery | *Clostridium_g4* | $1.275 \times 10^{-12}$ | 0 | $1.275 \times 10^{-12}$ |
| Family history | *Clostridium_g4* | 0 | $2.722 \times 10^{-12}$ | $2.722 \times 10^{-12}$ |
| Feeding | *Veillonella* | 0 | $1.010 \times 10^{-12}$ | $1.010 \times 10^{-12}$ |

**Table 3.** Summary statistics of clinical covariates included in the multi-omics analyses of data from the COCOA cohort. The p-values of the difference between two groups were obtained by chi-squared tests with the R package coin. AD, atopic dermatitis.

*Since only one sample had a family history score of 2, the corresponding sample was excluded from the analyses.

| Variable | Category | Controls (n = 46) | AD patients (n = 38) | P-value |
| --- | --- | --- | --- | --- |
| Sex | Male | 23 | 27 | 0.0504 |
|  | Female | 23 | 11 |  |
| Delivery mode | Caesarean | 15 | 11 | 0.718 |
|  | Vaginal | 31 | 27 |  |
| Feeding type | Breast | 13 | 15 | 0.552 |
|  | Mixed | 24 | 17 |  |
|  | Formula | 9 | 6 |  |
| Family history | 0 | 21 | 13 | 0.340 |
|  | 1 | 24 | 25 |  |
|  | 2 | 1* | 0 |  |

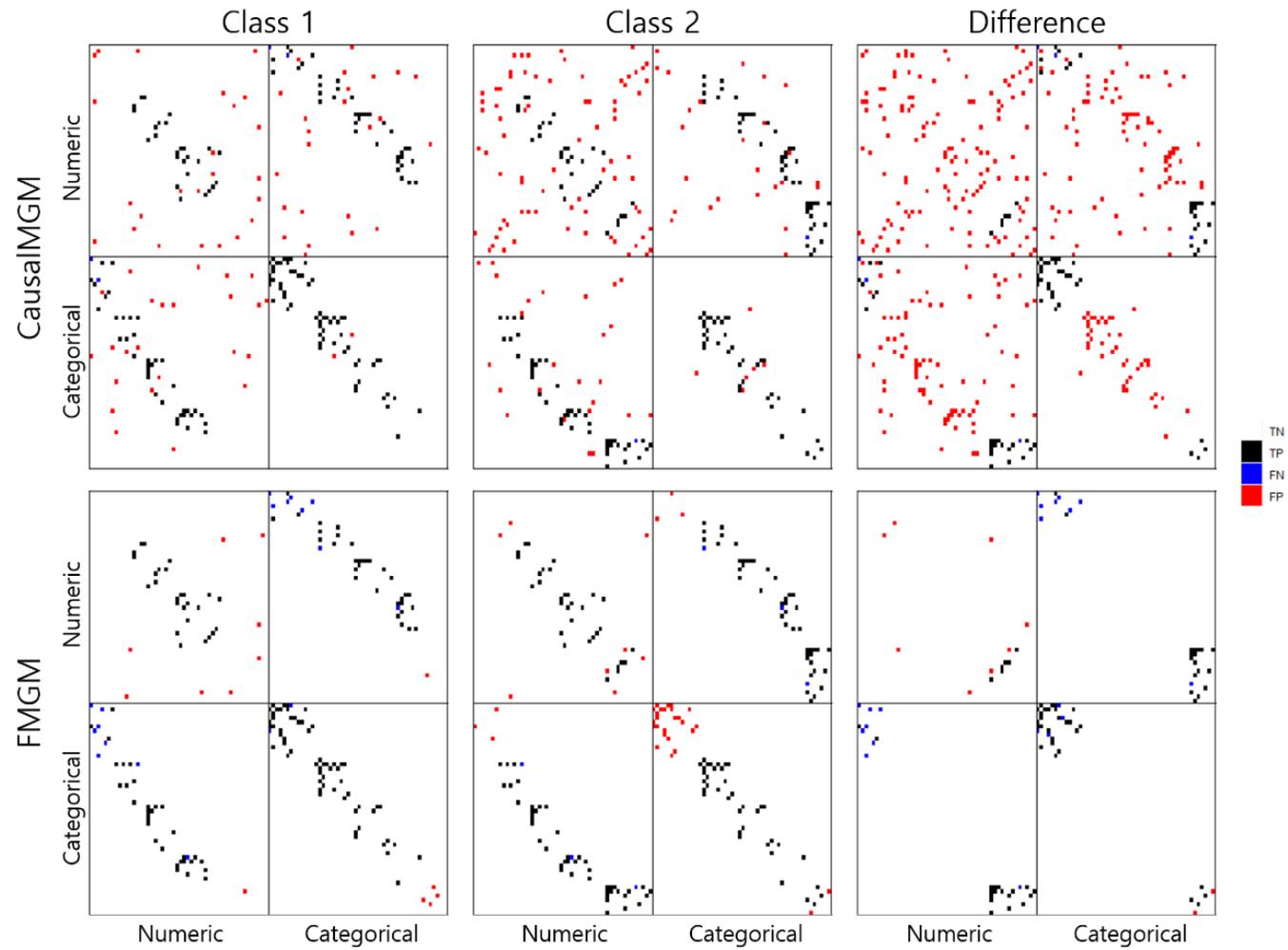

**Figure 1.** Overview of the inference results from simulated datasets, with causal FMGM for each class and fused MGM (FMGM). Inference results for each network and the difference are shown. Black dots represent true positives, while red and blue dots represent false positives and false negatives, respectively.

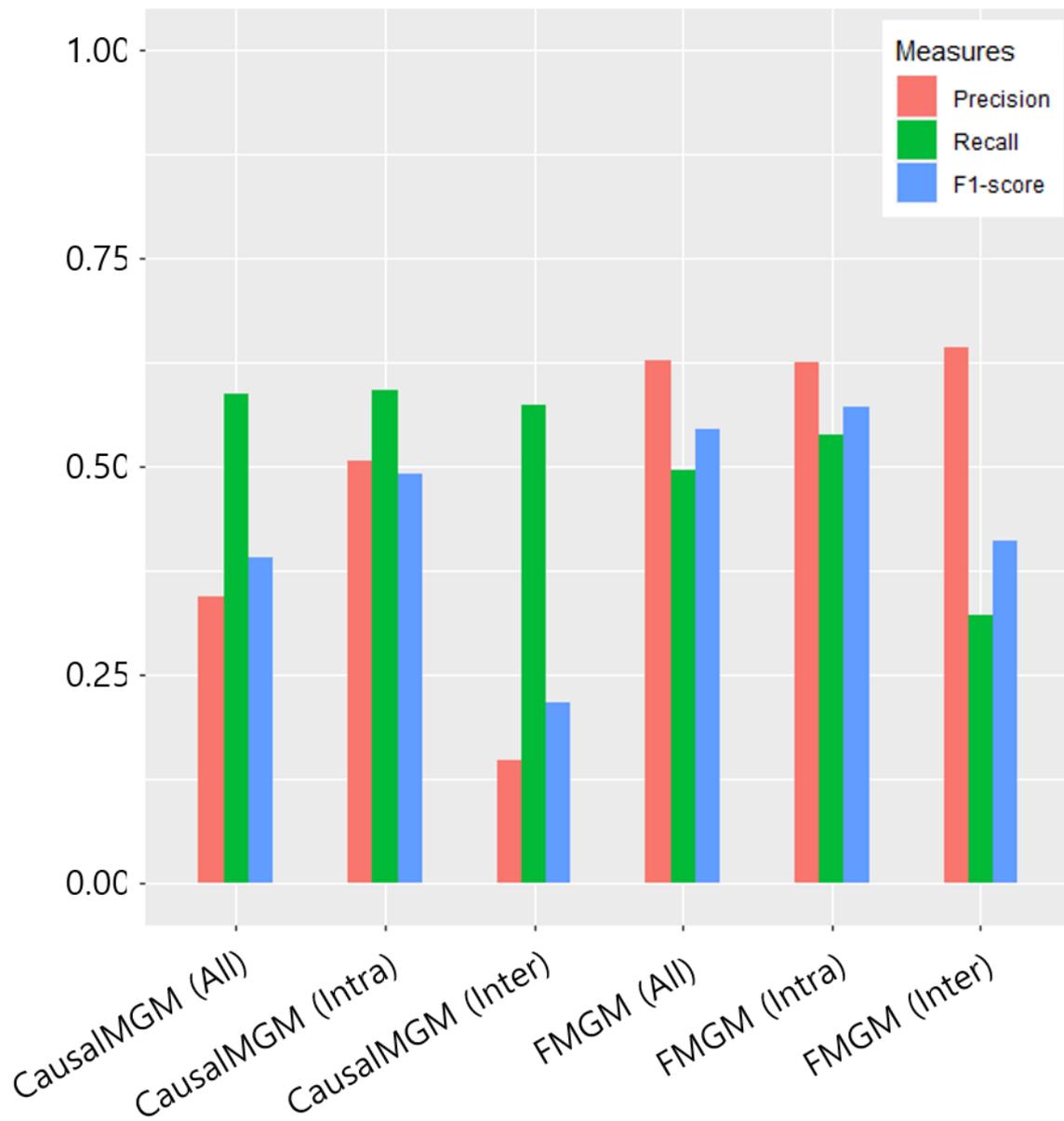

**Figure 2.** Overview of causal MGM (left) and fused MGM (FMGM, right) performance applied to simulated datasets. Precisions (red), recalls (green), and F1 scores (blue) are shown for overall inference (all), inference of networks (intra), and inference of differences (inter).

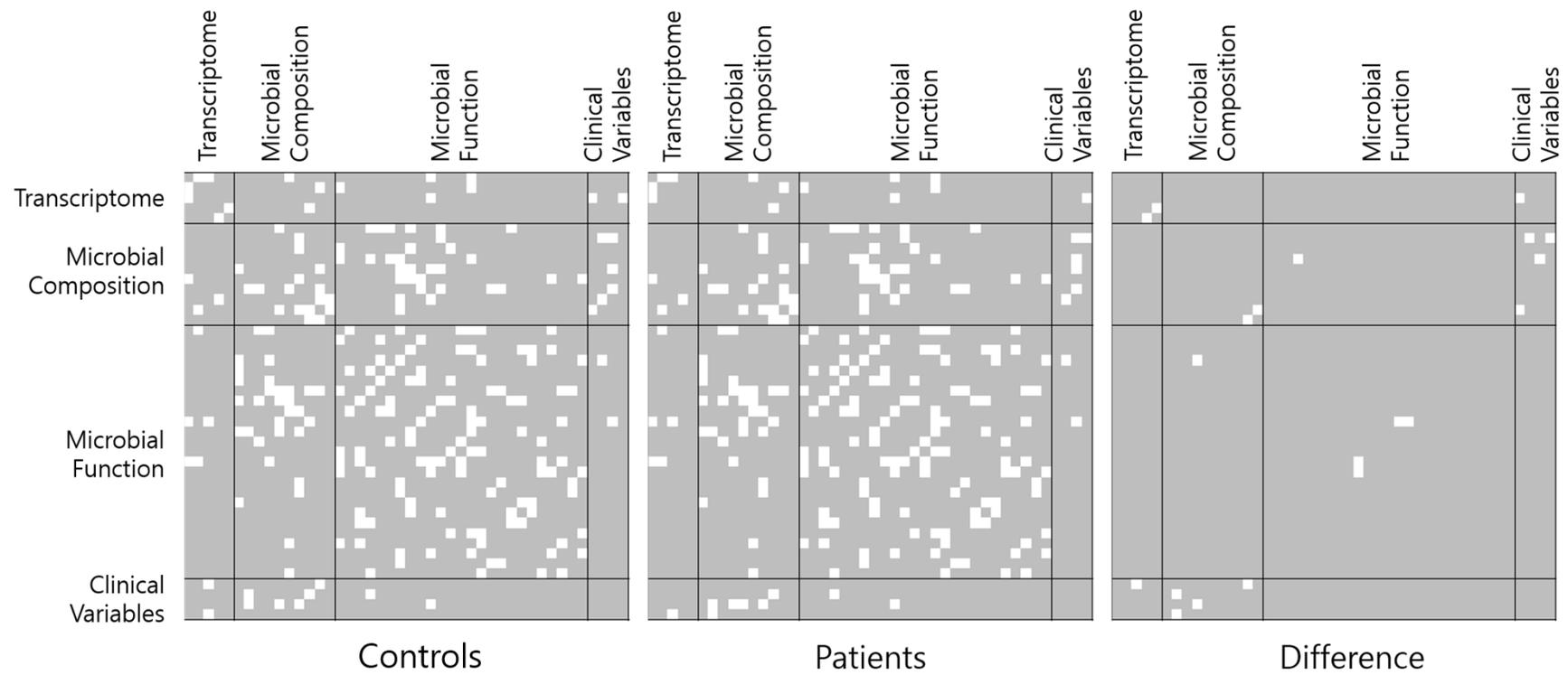

**Figure 3.** Overview of the network inference results underlying AD for patients and controls among 6-month-old infants, using multi-omics data and clinical variables. White dots indicate non-zero edges or non-zero differences.

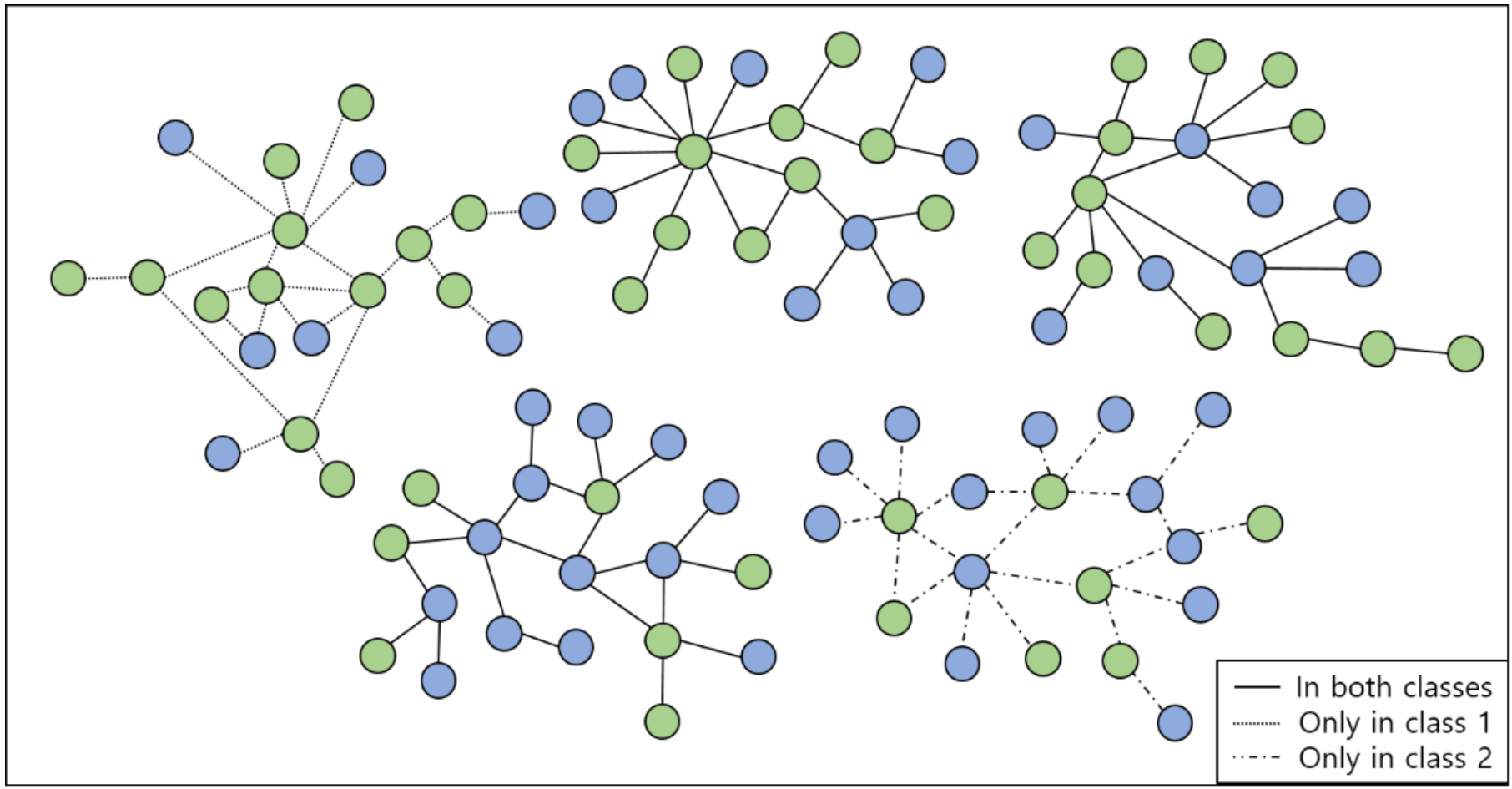

**Supplementary Figure 1**. An example of simulated scale-free network for simulation data analysis. One of five exclusively connected networks exists only in the first class, and another one exists only in the second class. Blue circles denote numeric variables, and green circles mean categorical variables.